\documentstyle[11pt,amsfonts]{article}

\newcommand{\be}{\begin{equation}}
\newcommand{\ee}{\end{equation}}
\newcommand{\bea}{\begin{array}}
\newcommand{\ea}{\end{array}}
\newcommand{\beqa}{\begin{eqnarray}}
\newcommand{\eeqa}{\end{eqnarray}}
\newcommand{\bean}{\begin{eqnarray*}}
\newcommand{\eean}{\end{eqnarray*}}

\def\up#1{\leavevmode \raise.16ex\hbox{#1}}

\def\sqr#1#2{{\vcenter{\vbox{\hrule height.#2pt
        \hbox{\vrule width.#2pt height#1pt \kern#1pt
          \vrule width.#2pt}
        \hrule height.#2pt}}}}

\def\BI{{\rm 1\!l}}

\newcommand{\gapproxeq}{\lower .7ex\hbox{$\;\stackrel{\textstyle
>}{\sim}\;$}}
\newcommand{\lapproxeq}{\lower .7ex\hbox{$\;\stackrel{\textstyle
<}{\sim}\;$}}

\newcounter{appendice}

\def\thebibliography#1{{\bf REFERENCES\markboth
 {REFERENCES}{REFERENCES}}\list
 {[\arabic{enumi}]}{\settowidth\labelwidth{[#1]}\leftmargin\labelwidth
 \advance\leftmargin\labelsep
 \usecounter{enumi}}
 \def\newblock{\hskip .11em plus .33em minus -.07em}
 \sloppy
 \sfcode`\.=1000\relax}

\def\Aone{\matrix{{}\cr {\cal  A} \cr {}^1}}
\def\Btwo{\matrix{{}\cr {\cal B} \cr {}^2}}

\def\psone{\matrix{{}\cr \psi \cr {}^1}}
\def\pstwo{\matrix{{}\cr \psi \cr {}^2}}
\def\bpsone{\matrix{{}\cr \bar\psi \cr {}^1}}
\def\bpstwo{\matrix{{}\cr \bar\psi \cr {}^2}}
\def\BI{{\rm 1\!l}}

\begin{document}


\centerline{ \LARGE  W-infinity Algebras from  Noncommutative  Chern-Simons Theory} 

\vskip 2cm

\centerline{ {\sc A. Pinzul and A. Stern}  }

\vskip 1cm
\begin{center}
Department of Physics, University of Alabama,\\
Tuscaloosa, Alabama 35487, USA
\end{center}

\vskip 2cm

\vspace*{5mm}

\normalsize
\centerline{\bf ABSTRACT}
We examine   Chern-Simons 
theory written  on a noncommutative plane with a `hole', and show that the algebra of observables is a
nonlinear deformation of the $w_\infty$ algebra.  The deformation
depends on the level
(the coefficient in the Chern-Simons action), and the noncommutativity
parameter, which were identified, respectively,
 with the inverse filling fraction (minus one) and   the inverse density in a recent description of
  the fractional quantum
Hall effect.   We
remark on the quantization of our algebra.  The results are sensitive
to the choice of ordering in the Gauss law.

\vspace*{5mm}

\newpage
\scrollmode

 An effective hydrodynamic description  of the fractional quantum Hall effect (FQHE)
in terms of noncommutative Chern-Simons theory was recently proposed
in \cite{sus}.   
It connected the  area  preserving diffeomorphism  symmetry of an
incompressible Hall fluid and with that present in
first order  noncommutative    Chern-Simons theory.   The symmetry is
generated by  the $w_\infty$ algebra, and it therefore should be
present in both contexts.  Within the context of Hall fluid, the role of  the $w_\infty$
 algebra and its quantization to $W_\infty$ (or
 $W_{1+\infty}$) algebras\cite{Shen} 
 has been discussed in a number of papers.\cite{IKS},\cite{ctz},\cite{Martst},\cite{Huerta}  Here
 we show how to recover the $w_\infty$ algebra (and 
 deformations thereof) from  noncommutative    Chern-Simons theory. 

It is well known how to write    Chern-Simons theory  on the
noncommutative plane $\times$ time\cite{ncs},  but  as with  the
commutative version, the theory is  empty.  This can be rectified with the introduction of sources.  In the commutative
  theory, sources can be introduced by punching holes in the plane.
  The noncommutative analogue of a punctured plane was developed in
\cite{nnsps}.  A `hole' was introduced by removing low lying states
from the Hilbert space.  Derivations could be  defined, and by
utilizing deformed coherent states\cite{mmsz},\cite{gps}, it was shown
how to recover the punctured plane in the commutative limit.  Here we write down 
Chern-Simons theory on such noncommutative spaces.  Imposing the necessary boundary
 conditions on the fields at the `hole', we find the resulting gauge
 invariant observables on phase space and show that their Poisson
 bracket algebra is a two parameter
nonlinear deformation of the $w_\infty$ algebra.  The two parameters are
 the `level' $k$ (having  integer values $\times \hbar$) and the noncommutativity parameter $\Theta_0$.  In
 \cite{sus},\cite{poly} the  integer values were identified with  inverse filling fractions
 $1/\nu$  (minus one) , and  $\Theta_0$ was identified with the inverse density (in
 the co-moving frame).  We find that the limit $k\rightarrow \infty$ gives the
 contraction to linear deformations $W_\infty$, while
 $\Theta_0\rightarrow 0$
 gives the  contraction to $w_\infty$.   We thereby recover the
 symmetry algebra for first order  noncommutative    Chern-Simons
 theory.  Our results are  in contrast to
 previous  quantum mechanical descriptions of the FQHE in terms of the
$linear$ deformations  $W_\infty$ (or $W_{1+\infty}$)
\cite{IKS},\cite{ctz},\cite{Martst},\cite{Huerta}.  At the end of this letter, we
remark on the quantization of our algebra, where we thus introduce a
third deformation parameter $\gamma $, which contains quantum
corrections and is sensitive to the choice of ordering in the Gauss law.  Initial results suggest
that $\gamma$ appears as an overall factor.

We first consider Chern-Simons theory written on the  noncommutative plane 
$\times$ time, which we denote by  ${\cal M}^{(0)}_F\times
{\mathbb{R}} $.   The noncommutative space ${\cal M}^{(0)}_F$ is
 generated by some  operator
$z$ and its  hermitian conjugate  $\bar z$, satisfying $[z,\bar
z]=\Theta_0$, $\Theta_0$ being a c-number.   $z$ and $\bar z$ have 
infinite dimensional representations.  We denote the vector space on
which they act by $
H^{(0)}$ ,   with basis vectors  $|n>\in H^{(0)} \;,n=0,1,2,...\;$.
The space   ${\cal M}^{(0)}_F$ admits
 derivations.  Derivatives on  ${\cal M}^{(0)}_F$ will be denoted by
 $\Delta$ and $\bar \Delta$, and assumed to commute
 \be [ \Delta, {\bar  \Delta}]=0
 \;,\label{qcdwd}\ee    Acting on a 
function $\Phi$ of $z$ and $\bar z$, 
\be { \Delta}\Phi=-i[{ p},\Phi]\;,\;\qquad {\bar \Delta}\Phi=-i[{ \bar
  p},\Phi]\;.\label{doph}\ee    The
 operator $p$ can be taken to be $-i\Theta_0^{-1}\bar z$, with
 its hermitian conjugate
 $\bar p=i\Theta_0^{-1}z$.  Then
  \be[ { p},{\bar  p}]=-\Theta^{-1}_0\;,\label{qqbar}\ee
which is consistent with  (\ref{qcdwd}).

The  degrees of freedom for noncommutative  Chern-Simons theory can be
taken to be a conjugate pair of potentials  $A$ and $\bar A$.   They are
functions on  ${\cal M}^{(0)}_F\times {\mathbb{R}} $.
Under gauge transformations:
\beqa  A&\rightarrow & i U^{\dagger} \Delta U+ U^{\dagger} AU \cr
 \bar A&\rightarrow & iU^{\dagger} \bar \Delta U + U^{\dagger} \bar
A U\;,\eeqa where $U$ is unitary
 function. 
  It is convenient to introduce 
$X=p+A$ and $\bar X =\bar p +\bar A$ for they   transform covariantly:
$ X\rightarrow U^{\dagger} XU\; $, $\; \bar X\rightarrow U^{\dagger} \bar
X U\;.$ 
  The  field strength is
\beqa { F}&=& i\Delta\bar A - i\bar\Delta A +[A,\bar A]\cr &=& [X,\bar X] -[p,\bar p]\;, \label{nccurv}
\eeqa
which then also transforms covariantly.  
  The  Chern-Simons Lagrangian can be written
 \beqa  L_{cs}&=&  k\Theta_0  \;{\rm Tr}\;\biggl( \frac i2(\dot A\bar
 A-A\dot{\bar A} ) + A_0 F \biggr) \cr & &\cr &=&
 k \;{\rm Tr}\;\biggl( \frac i2\Theta_0\;(D_{t}
X\bar X-X\overline{D_{t} X} )\;
 +\; A_0 \biggr) \;,\label{csl}\eeqa
where  
$ D_t X=\dot X -i[A_0,X]\;, $ the dot denotes a time derivative and
Tr is the trace  over basis states in $ H^{(0)}$.  We have dropped
total time derivatives and
equated terms related by cyclic permutation in going from the
first line to the second in (\ref{csl}).
  $A_0$ plays the role of a Lagrange
multiplier.  It is assumed to be hermitian and gauge transform as $ A_0  \rightarrow i U^\dagger
\dot U + U^\dagger A_0 U\;,
$ and so $ D_t X$ and its hermitian conjugate $\overline{D_{t} X}$
transform covariantly.

For gauge invariance one can
assume that $U$ and $U^\dagger $ act as the identity on 
$|n>$   as $n\rightarrow \infty$.  This corresponds to the requirement
in the commutative theory that gauge transformations vanish at spatial
infinity.  Applying the cyclic property of the trace, Tr$D_{t}
X\bar X$ and Tr$X\overline{D_{t} X} $ are  gauge invariant.
Concerning  the remaining term in (\ref{csl}),  the condition of invariance  was shown to lead to level
quantization.\cite{nair}, \cite{Bak} 
More precisely,  level
quantization follows from the demand that $\exp{ i\int_{ \mathbb{R} }
  dt\;L_{cs}} $ is invariant under gauge transformations satisfying
\be U \rightarrow \BI \;,\quad {\rm as}\quad t\rightarrow \pm \infty
\label{uastgi}\ee  The quantization condition is $ k= {\rm
  integer}\times \hbar$, and the integer
was identified in \cite{sus},\cite{poly} with the
inverse of the filling fraction $\nu$ (minus one) .

As with Chern-Simons theory on commutative ${\mathbb{R}}^3$, the above
 theory is empty.  This is easily seen in the canonical formalism. 
 The time derivative terms  in (\ref{csl}) define the Poisson
 structure.   The phase space is spanned by matrix elements $
\chi_n^{\;m} = <n|X|m > $ and $  \bar\chi_n^{\;m} = <n|\bar X|m>  $,
 with Poisson brackets
\be \{ 
\chi_n^{\;m} , \bar \chi_r^{\;s} \} =-\frac{i}{k\Theta_0}  \delta^m_r
 \delta^s_n \ee
 The remaining terms  in the trace in (\ref{csl}) give the Gauss law constraints
\be  G_n^{\;m} = <n|[X,\bar X]|m > +\Theta_0^{-1}\delta_n^m  = \chi_n^{\;r} \bar \chi_r^{\;m} - \bar  \chi_n^{\;r}  \chi_r^{\;m}
 +\Theta_0^{-1} \delta_n^m \approx 0 \;. \ee  They are first class, and  from
 $$ ik\Theta_0 \;\{ 
\chi_n^{\;m}, G_r^{\;s}\} = \chi_r^{\;m}\delta_n^s -  \chi_n^{\;s}\delta_r^m $$
 $$ ik\Theta_0 \; \{ \bar
\chi_n^{\;m}, G_r^{\;s}\} =\bar \chi_r^{\;m}\delta_n^s - \bar
\chi_n^{\;s}\delta_r^m $$  generate gauge transformations.  Since
every first class constraint eliminates two   phase space variables, no degrees of freedom remain after projecting to the
reduced phase space.
     
Alternative noncommutative spaces were examined in \cite{nnsps}.  We
denote them by  ${\cal M}^{(n_0)}_F$.
Their commutative limit was shown to be the punctured plane with a
nontrivial Poisson structure.     For convenience we again
call  the
  generators of the algebra  by $z$ and $\bar z$, although  now  $[z,\bar
z]\ne\Theta_0$.  These generators
  act on a  Hilbert space $ H^{(n_0)}$ which is an
 infinite dimensional subset of $  H^{(0)} $.    $ H^{(n_0)}$ is spanned
 by basis vectors  $|n>, \; n=n_0,n_0+1,n_0+2,...$, $n_0$ being some
positive integer.  We have thus put a `hole' in the Hilbert space  $
H^{(0)} $.  ${\cal M}^{(n_0)}_F$  was  shown to admit
 derivations.  We once again denote derivatives by
 $\Delta$ and $\bar \Delta$, and assume they commute.  Now introduce a
 function $\Phi$ on  ${\cal M}^{(n_0)}_F$.  It can be
 nonvanishing only on  vectors belonging to   $ H^{(n_0)}$.  As before,
 we assume (\ref{doph}),  so we   need (\ref{qqbar}).  An explicit expression for
 $p$ and $\bar p$ in terms of $z$ and $\bar z$ was given in
 \cite{nnsps}, but it is not necessary here.   (\ref{qqbar}) shows
 that  $p$ and $\bar p$
 are proportional to the usual raising and lowering operators,
 respectively.  (We take $\Theta_0>0$.)  But then $\bar p$ takes vectors out of
 $ H^{(n_0)}$, while $p$ takes (bra) vectors out of the dual space.
 For derivations to be well defined we then must  impose `boundary
 conditions' on fields at the `hole'.    ${\bar \Delta}\Phi$ is well defined
 on  $ H^{(n_0)}$ when $<n_0|\Phi|n> =0$,  while   ${ \Delta}\Phi$ is well defined
 on  the dual space when $<n|\Phi|n_0> =0$.  Stronger boundary
 conditions are needed for higher derivatives  to be defined.  On the
 other hand, for Chern-Simons theory one only needs first order
 derivatives.\footnote{We regard $A$ and $\bar A$ - and  not $X$
   and $\bar X$ - as the fundamental
configuration space variables.}  More specifically, the derivatives 
$\Delta\bar A$ and $\bar\Delta A$ should be well defined as they
 appear in the field strength (\ref{nccurv}), and so our boundary
 conditions are:
\be <n_0|A|n>=<n|\bar A|n_0> =0\;,\qquad \forall\; n\ge n_0\ee
Since $p$ and $\bar p$
 are proportional to raising and lowering operators, respectively, we can also write
\be <n_0|X|n>=<n|\bar X|n_0> =0\;,\qquad \forall\; n\ge n_0\label{bcox}\ee
In order that these boundary conditions are preserved under gauge
 transformations we need the unitary matrices to satisfy
\be <n_0|U^\dagger |a>=<a|U|n_0>=0 \;,\qquad  \forall\;a,b,...\ge n_0+1\label{rou} \ee
Since gauge transformations are thereby restricted, not all phase
 space degrees of freedom in Chern-Simons theory can be gauged away,
 as was the case previously.

For the Chern-Simons Lagrangian we once again assume  (\ref{csl}),
 only now the trace is over a basis in
   $ H^{(n_0)}$.
Returning to the Hamiltonian formulation, and now imposing the
 boundary conditions (\ref{bcox}), one is left with  the following  phase space variables:
 $$
\chi_a^{\;b} =\Theta_0 <a|X|b >   \qquad  \bar\chi_a^{\;b} =\Theta_0 <a|\bar X|b> \;, $$
 $$ 
\psi_a = <a| X|n_0 >  \qquad  \bar  \psi^a = <n_0|\bar X|a>\;, $$
where again $a,b,...>n_0$, and we  have rescaled $\chi$ and $\bar\chi$
in order to later obtain the desired commutative limit. The nonzero Poisson brackets are
\be \{ 
\chi_a^{\;b} , \bar \chi_c^{\;d} \} =-\frac{i}{k}{\Theta_0}\; \delta^b_c \delta^d_a \qquad
 \{ 
\psi_a , \bar \psi^b \} = -\frac{i}{k\Theta_0} \delta^b_a \label{pbcps} \ee
For later convenience we also re-scale the Gauss law
constraints:\footnote{For simplicity, we shall assume that there are
  no further
  constraints $ G_{n_0}^{\;\;a}\approx  G_a^{\;n_0}\approx G_{n_0}^{\;\;n_0}\approx 0$.  For this we may set the $n_0^{th}$ row and column of
  the Lagrange multiplier $A_0$
  equal to zero.} 
\be G_a^{\;b} =\Theta_0^2 <a|[X,\bar X]|b > +\Theta_0\delta_a^b  = \chi_a^{\;c} \bar \chi_c^{\;b} - \bar  \chi_a^{\;c}  \chi_c^{\;b}+\Theta_0\delta_a^b
+\Theta_0^2\psi_a  \bar \psi^b  \approx 0
\label{tglc}\ee
They generate gauge transformations which are consistent with (\ref{rou}):
 \beqa ik\Theta_0^{-1} \; \{ 
\chi_a^{\;b}, G_c^{\;d}\} &=& \chi_c^{\;b}\delta_a^d -  \chi_a^{\;d}\delta_c^b \cr
  ik\Theta_0^{-1} \; \{ \bar
\chi_a^{\;b}, G_c^{\;d}\}& =&\bar \chi_c^{\;b}\delta_a^d - \bar
\chi_a^{\;d}\delta_c^b \cr ik\Theta_0^{-1} \; \{ 
 \psi_a, G_b^{\;c}\} &=&     \psi_b\delta_a^c \cr
   ik\Theta_0^{-1} \; \{ \bar
 \psi^a, G_b^{\;c}\}&= &-  \bar   \psi^c\delta_b^a \eeqa
From a counting argument alone the variables $
\chi_a^{\;b}$ and $  \bar\chi_a^{\;b}$ can be gauged away, leaving only
$ 
\psi_a $ and $  \bar  \psi^a$.  But the latter are not gauge
invariant.  Instead they transform as a vector and conjugate vector,
while  $
\chi_a^{\;b}$ and $  \bar\chi_a^{\;b}$ transform as tensors.  Then we can  construct
gauge invariant  observables of the form $ \bar  \psi {\cal A}\psi\;, $
where   ${\cal A}$ denotes
 polynomial functions  in the fields $\chi$ and $\bar\chi$.  It
 remains to   compute their Poisson bracket algebra.  For this
we can use \be \{ \bar  \psi {\cal A}\psi , \bar  \psi {\cal B}\psi \}
 = -\frac{i}{k\Theta_0}\bar\psi [{\cal A},{\cal B}] \psi + \bpsone\bpstwo \{\Aone,\Btwo \}
\psone\pstwo \;, \label{pabpbp}\ee where $[\;,\;]$ is the commutator bracket.  The labels $1$ and $2$ indicate  two separate  vector
spaces, where for example $\Aone$ and $\Btwo$ are the tensor products
$ {\cal A}\otimes \BI$ and $\BI\otimes {\cal  B} ,$ respectively,
$\BI$ being the unit operator.  If we denote the right hand side of (\ref{pabpbp})
by $\bar \psi{\cal O}_{{\cal A},{\cal B}}\psi=-\bar \psi{\cal O}_{{\cal B},{\cal A}}\psi $, then 
$$ \{ \bar  \psi {\cal A}\psi , \bar  \psi {\cal BC}\psi \} =
\bar\psi{\cal O}_{{\cal A},{\cal B}}{\cal C}\psi
+\bar\psi {\cal B}{\cal O}_{{\cal A},{\cal C}}\psi  $$ 
We note that ${\cal O}_{{\cal A},{\cal B}}$ can depend on $\psi$ and $\bar\psi$, so
that  $ \bar  \psi {\cal A}\psi $ do not generate a linear algebra.  
Furthermore, from the Gauss law
constraint (\ref{tglc}),   observables obtained via a reordering of the
$\chi$ and $\bar \chi$ factors in ${\cal A}$ form an equivalence class.  We
fix a gauge by choosing the following ordering 
\be M_{(\alpha,\beta)} =- k \;\bar \psi (\bar \chi)^\alpha
(\chi)^\beta\psi  \label{defofM} \ee
  From (\ref{pabpbp}),  $ M_{(0,0)}  $ is a central charge. 
 Examples of   nonzero Poisson brackets are:
\beqa 
\{ M_{(0,1)}, M_{(1,0)} \}& =& -iM_{(0,0)}  \cr
 \{ M_{(0,1)}, M_{(1,1)} \}& =&-i M_{(0,1)}   \cr
 \{ M_{(1,1)}, M_{(1,0)} \} &=& -iM_{(1,0)}   \cr 
 \{ M_{(0,1)}, M_{(2,0)} \} &=&- 2i M_{(1,0)} \cr
 \{ M_{(0,2)}, M_{(1,0)} \}&=& -2iM_{(0,1)}   \cr
 \{ M_{(0,2)}, M_{(1,1)} \}& =& -2iM_{(0,2)}  \cr
\{ M_{(1,1)}, M_{(2,0)} \}& =&- 2i M_{(2,0)}  \cr
 \{ M_{(0,2)}, M_{(2,0)} \}& =&- 4i M_{(1,1)}  +  2i\Theta_0
 M_{(0,0)} - \frac{2i}k \Theta_0^2 M_{(0,0)}^2 \label{espbs}
\eeqa
where we used the Gauss law constraint (\ref{tglc}) to do reordering. The  last example shows
that the algebra is nonlinear.

Although we don't have a closed form expression for the algebra, there
 are some familiar contractions.  The commutative limit is $\Theta_0\rightarrow 0 $.   Both the
 Poisson bracket and the commutator of any  two polynomials ${\cal A}$
 and ${\cal B}$ of
$\chi$ and $\bar \chi$  are linear in $\Theta_0$ to leading order.
 For the former the result follows
from (\ref{pbcps}), while for the latter the result follows
from  the Gauss law constraint  (\ref{tglc}).  Then the second term in
(\ref{pabpbp}) can be dropped.  Moreover, at the lowest nontrivial order we can
 represent $\chi$ and
 $\bar\chi$ by commuting numbers $\zeta$ and $\bar \zeta$, respectively, and replace the commutator bracket $[\;,\;]$ by $\widetilde{\{\;,\;\}}$,  with
 $$\widetilde{\{{\cal A},{\cal B}\}}=\Theta_0\biggl(\frac{\partial
 {\cal A}}{\partial  \zeta}\frac{\partial {\cal B}}{\partial\bar \zeta }
-\frac{\partial {\cal A}}{\partial \bar
 \zeta}\frac{\partial {\cal  B}}{\partial \zeta}\biggr)$$   Then in the limit
\beqa \{ M_{(\alpha,\beta)}, M_{(\rho,\sigma)} \}&\rightarrow &
 -ik\Theta_0^{-1}\;\bar\psi \; \widetilde{\{\bar \zeta^\alpha \zeta^\beta,\bar
 \zeta^\rho \zeta^\sigma\}}\; \psi \cr
 & & =-i (\beta\rho -\alpha
 \sigma)\;
M_{(\alpha+\rho - 1, \beta + \sigma -1)}  \eeqa
This is the  `classical' $w_\infty$ algebra which is associated with  area preserving
 diffeomorphisms.  (Actually, as in \cite{ctz} we get only a subalegra
 of the
 $w_\infty$ algebra since no negative values for $\alpha,\beta,...$
 are allowed,  restricting to the nonsingular  area preserving
 diffeomorphisms of the plane.)  On the other hand, away from the
limit of vanishing $\Theta_0$ we get a deformation of
 the  $w_\infty$ algebra.  It is in fact a two-parameter  nonlinear deformation, the other
 parameter being the level $k$, which parameterizes the  nonlinearity. From  (\ref{pbcps}), any $n$-th order term in the Poisson bracket algebra
 goes like $k^{1-n}$.  Since $k$ is identified with the inverse of the filling fraction
 $\nu$ in the FQHE, we get a different algebra for different values of $\nu$. 
 The nonlinear terms tend to zero for large $k$ ( small 
 $\nu$) and  then we approach the linear `quantum' $W_\infty$
 algebra.\footnote{More accurately,
 due to the absence of negative values for $\alpha,\beta,...,$ we approach a
 subalgebra of $W_{1+\infty}$, which includes
 the so-called `wedge' algebra $W_\Lambda$ \cite{ctz}.} 

 The role of  linear deformations of the $w_\infty$
 algebra in the quantum mechanical description of the FQHE 
 has been discussed in a number of
 papers.\cite{IKS},\cite{ctz},\cite{Martst}  They make up the edge variables
for the system on a finite size domain.  In contrast, for arbitrary $k$ and $\Theta_0$, we have obtained a nonlinear deformation of  the
 $w_\infty$ algebra in the candidate effective theory 
 for the FQHE.  (Nonlinear deformations of the  $w_\infty$ algebra were
  obtained previously in  different contexts \cite{nldfs},  
and from  \cite{yuwu} such deformations are unique.)  A further distinction is that our deformation appears already at the $classical$ $level$ of 
 noncommutative Chern-Simons theory.  It is  not clear whether  $ M_{(\alpha,\beta)}$
 are `edge' variables.   Presumably  they live in the
 vicinity of the `puncture' in the commutative limit.  A careful analysis of
 the continuum limit is required to verify this. 
 
  There has been recent work on finite
dimensional  matrix models with the hope of describing a quantum Hall
droplet.\cite{poly},\cite{hol},\cite{lug}  In this regard, instead of working with the infinite dimensional Hilbert space  $
H^{(n_0)}$, as we did above, we can repeat the analysis for its complement $\bar
H^{(n_0)}$, which is  finite dimensional.    $\bar H^{(n_0)}$ is spanned
 by basis vectors  $|n>,$  $ n=0,1,2,...,n_0-1$.    We  denote the
 corresponding noncommuting space by   $\bar{\cal M}^{(n_0)}_F$, and
 its derivatives once again by
 $\Delta$ and $\bar \Delta$, which are assumed to commute.  A
 function $\Phi$ on $\bar{\cal M}^{(n_0)}_F$ is defined to be nonvanishing only on    $\bar
 H^{(n_0)}$.  Acting on 
  $\Phi$,  $\Delta$ and $\bar \Delta$ are  written as in (\ref{doph}).  This implies (\ref{qqbar}), so again   $p$ and $\bar p$
 are proportional to the usual raising and lowering operators,
 respectively.  Now  $ p$ takes vectors out of
 $\bar H^{(n_0)}$, while $\bar p$ takes (bra) vectors out of the dual space.
 The necessary boundary
 conditions are    $<n_0-1|\Phi|n> =0$ for    ${ \Delta}\Phi$ to
 be well defined, and $<n|\Phi|n_0-1> =0$ for   ${\bar \Delta}\Phi$ to
 be well defined.  Then for Chern-Simons theory, one has:
$ <n_0-1|\bar A|n>=<n| A|n_0-1> =0\;, \forall\; n\le n_0-1$, with
analogous conditions on the matrix elements of $X$ and $\bar X$.  In order that these boundary conditions are preserved under gauge
 transformations:  $ <n_0-1|U^\dagger |a>=<a|U|n_0-1>=0 ,$ where here
$ a,b,...=0,1,2,..., n_0-2 $.  Gauge transformations are generated by
(\ref{tglc}), with $ a,b,...=0,1,2,..., n_0-2 $.  Now all but $2(n_0 -1)$
 phase
 space variables can be gauged away.  The remaining gauge invariant
 variables can again be written as
(\ref{defofM}) (where $ 
\psi_a = <a| X|n_0-1 >$ and $  \bar  \psi^a = <n_0-1|\bar X|a>\; $),
although for finite $n_0$ they are not all independent degrees of
freedom.  For example, the trace of (\ref{tglc}) gives $\psi_a  \bar \psi^a = (n_0 -1) \Theta_0^{-1} $.  In the
commutative limit we should let $n_0\rightarrow \infty$ (in addition
to $\Theta_0 \rightarrow 0$), so we
again recover the $w_\infty$ algebra.

The phase space description of the finite dimensional  system described above is in agreement with
that of Polychronakos\cite{poly}.  Although we   don't
introduce  vector degrees of freedom in the Lagrangian as in
\cite{poly},  analogous phase space degrees of freedom $\psi_a$ and
$\bar\psi^a$ appear  at the Hamiltonian level.   A potential term is
introduced in \cite{poly} and the resulting dyncamics is  claimed to be
equivalent the Calogero system.

Additional deformations of our nonlinear $W_\infty$ algebra can occur
after quantization, with the possible inclusion of central terms.
One  quantization program is to replace the original phase variables
$
\chi_a^{\;b}$, $ \bar\chi_a^{\;b}$, $\psi_a$, $  \bar  \psi^a$ by
the quantum operators $
\hat\chi_a^{\;b}$, $\hat{ \bar\chi}_a^{\;b}$, $\hat\psi_a$, $ \hat{ \bar  \psi}^a$,
respectively, and  Poisson brackets  (\ref{pbcps}) by the commutation relations: 
\be [ 
\hat\chi_a^{\;b} , \hat{\bar \chi}_c^{\;d}] =\frac{\hbar\Theta_0} {k}\; \delta^b_c \delta^d_a \qquad
[\hat
\psi_a ,\hat {\bar \psi}^b ] = \frac{\hbar}{k\Theta_0} \delta^b_a \;,
\label{crcps} \ee thereby introducing the additional deformation parameter
$\hbar$.  Next we choose the following ordering for the Gauss law  operators
\be \hat G_a^{\;b} = \hat\chi_a^{\;c} \hat{\bar \chi}_c^{\;b} -\hat{ \bar  \chi}_a^{\;c} \hat \chi_c^{\;b}+\Theta_0\delta_a^b
+\Theta_0^2\hat\psi_a \hat{ \bar \psi}^b 
\label{qtglc}\ee
It can be checked that their commutator algebra closes, and so we can consistently
impose that they vanish on physical states.  The operator analogues $\hat
 M_{(\alpha,\beta)}$ of the gauge invariant quantities (\ref{defofM})
 can be constructed, and their algebra computed.  As in the classical case,
 $\hat  M_{(0,0)}$ is central.   Upon computing
the quantum analogues of Poisson brackets  (\ref{espbs}), we get
\beqa 
[\hat M_{(0,1)},\hat M_{(1,0)}] & =& \gamma \hat M_{(0,0)}  \cr
[\hat M_{(0,1)},\hat M_{(1,1)}]& =&   \gamma\hat M_{(0,1)}   \cr
[\hat M_{(1,1)},\hat M_{(1,0)}] &=&   \gamma\hat M_{(1,0)}   \cr 
[\hat M_{(0,1)},\hat M_{(2,0)} ] &=& 2\gamma\hat M_{(1,0)} \cr
[\hat M_{(0,2)},\hat M_{(1,0)} ]&=& 2 \gamma\hat M_{(0,1)}   \cr
[\hat M_{(0,2)},\hat M_{(1,1)}]& =& 2 \gamma\hat M_{(0,2)}  \cr
[\hat M_{(1,1)},\hat M_{(2,0)} ]& =&2 \gamma\hat M_{(2,0)}  \cr
[\hat M_{(0,2)},\hat M_{(2,0)}]& =& \gamma\biggl(4 \hat M_{(1,1)}  -  2\Theta_0
\hat M_{(0,0)} + \frac{2}k \Theta_0^2\hat M_{(0,0)}^2\biggr) \label{escrs}
\eeqa  where the $\gamma$ factor contains $\hbar$ corrections.     From  (\ref{escrs}) it appears that
$\gamma$ may be   an overall factor in the  quantum
commutators.  For the
 choice of 
 ordering in (\ref{qtglc}), one gets
$\gamma=\hbar (1+ \hbar/ k) $.  On the other hand, if  $\hat\psi_a$ and $ \hat{ \bar \psi}^b $ are switched in
 the last term of  (\ref{qtglc}), then $\gamma=\hbar$.    $\gamma$ can
 be re-expressed in terms of  the filling fraction $\nu$.  According to \cite{poly}, $\nu^{-1} =1+  k/\hbar$.

The  task of writing down a closed form expression for the full quantum algebra
appears to be   nontrivial.
After  this hurdle, one is next faced with the task of  finding unitary representations.  Although representation theory for
linear deformations of  the
 $w_\infty$ algebra  is known\cite{KacRad}, the same cannot be said for the  nonlinear
 deformations.  If the quantization program   can be
 successfully carried out it should offer a nice  test for the noncommutative
 Chern-Simons description of the FQHE.  Lastly, we remark  that the
 exhibition of  the noncommutative $W_\infty$ algebra could be
 helpful in  recovering the commutative limit.  In that limit, we
 should somehow recover Chern-Simons theory on a  domain with a boundary.  (An
 attempt along these lines was made  in \cite{hol}.)   The latter is known to
 have  all its degrees of freedom  at the spatial boundary.  These are
 the so called `edge states', which are  associated with a conformal algebra, or more
 generally a $w_\infty$ algebra.  Thus our gauge invariant observables
 should get  mapped to the edge states in the limit.

\end{document}